\documentclass[aps,prl,preprint,tightenlines,superscriptaddress,showpacs,byrevtex]{revtex4}
\usepackage{graphicx} 
\usepackage{dcolumn}  

\graphicspath{{ps}}

\begin{document}


\preprint{\vbox{ \hbox{   }
                 \hbox{BELLE-CONF-0673}
}}

\title{Studies of $\gamma\gamma\to\Lambda\bar{\Lambda},\Sigma^0\bar{\Sigma^0}$ 
production at Belle}


\affiliation{Budker Institute of Nuclear Physics, Novosibirsk}
\affiliation{Chiba University, Chiba}
\affiliation{Chonnam National University, Kwangju}
\affiliation{University of Cincinnati, Cincinnati, Ohio 45221}
\affiliation{University of Frankfurt, Frankfurt}
\affiliation{The Graduate University for Advanced Studies, Hayama} 
\affiliation{Gyeongsang National University, Chinju}
\affiliation{University of Hawaii, Honolulu, Hawaii 96822}
\affiliation{High Energy Accelerator Research Organization (KEK), Tsukuba}
\affiliation{Hiroshima Institute of Technology, Hiroshima}
\affiliation{University of Illinois at Urbana-Champaign, Urbana, Illinois 61801}
\affiliation{Institute of High Energy Physics, Chinese Academy of Sciences, Beijing}
\affiliation{Institute of High Energy Physics, Vienna}
\affiliation{Institute of High Energy Physics, Protvino}
\affiliation{Institute for Theoretical and Experimental Physics, Moscow}
\affiliation{J. Stefan Institute, Ljubljana}
\affiliation{Kanagawa University, Yokohama}
\affiliation{Korea University, Seoul}
\affiliation{Kyoto University, Kyoto}
\affiliation{Kyungpook National University, Taegu}
\affiliation{Swiss Federal Institute of Technology of Lausanne, EPFL, Lausanne}
\affiliation{University of Ljubljana, Ljubljana}
\affiliation{University of Maribor, Maribor}
\affiliation{University of Melbourne, Victoria}
\affiliation{Nagoya University, Nagoya}
\affiliation{Nara Women's University, Nara}
\affiliation{National Central University, Chung-li}
\affiliation{National United University, Miao Li}
\affiliation{Department of Physics, National Taiwan University, Taipei}
\affiliation{H. Niewodniczanski Institute of Nuclear Physics, Krakow}
\affiliation{Nippon Dental University, Niigata}
\affiliation{Niigata University, Niigata}
\affiliation{University of Nova Gorica, Nova Gorica}
\affiliation{Osaka City University, Osaka}
\affiliation{Osaka University, Osaka}
\affiliation{Panjab University, Chandigarh}
\affiliation{Peking University, Beijing}
\affiliation{University of Pittsburgh, Pittsburgh, Pennsylvania 15260}
\affiliation{Princeton University, Princeton, New Jersey 08544}
\affiliation{RIKEN BNL Research Center, Upton, New York 11973}
\affiliation{Saga University, Saga}
\affiliation{University of Science and Technology of China, Hefei}
\affiliation{Seoul National University, Seoul}
\affiliation{Shinshu University, Nagano}
\affiliation{Sungkyunkwan University, Suwon}
\affiliation{University of Sydney, Sydney NSW}
\affiliation{Tata Institute of Fundamental Research, Bombay}
\affiliation{Toho University, Funabashi}
\affiliation{Tohoku Gakuin University, Tagajo}
\affiliation{Tohoku University, Sendai}
\affiliation{Department of Physics, University of Tokyo, Tokyo}
\affiliation{Tokyo Institute of Technology, Tokyo}
\affiliation{Tokyo Metropolitan University, Tokyo}
\affiliation{Tokyo University of Agriculture and Technology, Tokyo}
\affiliation{Toyama National College of Maritime Technology, Toyama}
\affiliation{University of Tsukuba, Tsukuba}
\affiliation{Virginia Polytechnic Institute and State University, Blacksburg, Virginia 24061}
\affiliation{Yonsei University, Seoul}
  \author{K.~Abe}\affiliation{High Energy Accelerator Research Organization (KEK), Tsukuba} 
  \author{K.~Abe}\affiliation{Tohoku Gakuin University, Tagajo} 
  \author{I.~Adachi}\affiliation{High Energy Accelerator Research Organization (KEK), Tsukuba} 
  \author{H.~Aihara}\affiliation{Department of Physics, University of Tokyo, Tokyo} 
  \author{D.~Anipko}\affiliation{Budker Institute of Nuclear Physics, Novosibirsk} 
  \author{K.~Aoki}\affiliation{Nagoya University, Nagoya} 
  \author{T.~Arakawa}\affiliation{Niigata University, Niigata} 
  \author{K.~Arinstein}\affiliation{Budker Institute of Nuclear Physics, Novosibirsk} 
  \author{Y.~Asano}\affiliation{University of Tsukuba, Tsukuba} 
  \author{T.~Aso}\affiliation{Toyama National College of Maritime Technology, Toyama} 
% ...
\collaboration{Belle Collaboration}
\noaffiliation

\begin{abstract}

Cross sections for hyperon pair production from two-photon collisions, 
$\gamma\gamma\to\Lambda\bar{\Lambda},\Sigma^0\bar{\Sigma^0}$, are measured 
in the $2-4$ GeV energy region at Belle, using 464 fb$^{-1}$ of data. 
A contribution from the intermediate resonance $\eta_c(1S)$ is observed, 
and the products of the two-photon width of the $\eta_c(1S)$ and its branching 
ratios to $\Lambda\bar{\Lambda}$ and $\Sigma^0\bar{\Sigma^0}$ are measured. 
The results will help test QCD models.

\end{abstract}


\maketitle

\tighten

{\renewcommand{\thefootnote}{\fnsymbol{footnote}}}
\setcounter{footnote}{0}

\section{Introduction}

Measurement of hyperon-pair production from two-photon collisions is  
important for the study of QCD models and flavour symmetry.
Previous measurements \cite{CL,L3} have been performed with very 
limited statistics; the results of different experiments
are not consistent.

General theories of hard exclusive processes 
in QCD~\cite{cz1,bl,CZR}
and the pure-quark picture 
\cite{FA} 
result in an expected cross section for baryon-pair production from 
two-photons that is one-order of magnitude below the experimental data. 
In order to explain the experimental observation, 
various models were proposed. 
For example,
the diquark model \cite{MA,KP,CF5,CF6} 
achieves better agreement in the absolute size of
the cross section. 
In handbag approaches \cite{HB}, the process is factorized into a hard
$\gamma\gamma\to q\bar{q}$ sub-process and a soft $q\bar{q}\to$ baryon-pair 
transition; the expected 
cross section is determined
by the so-called \emph{annihilation form-factors}.
In these calculations,  
all effects from flavour symmetry breaking are neglected.
More recently, predictions were also provided within pole and resonance
approaches 
\cite{OD}.

Using 464 fb$^{-1}$ of data collected 
at Belle, hundreds of events have been obtained for 
each of the two channels
$\gamma\gamma\to\Lambda\bar{\Lambda},\Sigma^0\bar{\Sigma^0}$. 
Contributions from $\eta_c$ resonances are also observed.
The measured cross sections will help to
test all the existing models more precisely. 

\section{Belle detector and event selection}

Data are recorded with the Belle detector \cite{belle} at
KEKB \cite{kekb}, which is an asymmetric $e^+e^-$ collider 
operated at 10.58~GeV center-of-mass (c.m.) energy. 
The following Belle subsystems are of importance 
for our analyses: the central drift chamber (CDC), the aerogel Cherenkov 
counters (ACC), the time-of-flight scintillation counters (TOF) and the 
CsI($Tl$) electromagnetic calorimeter (ECL). 
The CDC measures the momenta of charged particles and provides
precise ($6\%$) $dE/dx$ measurements. 
The ACC measures the number of photoelectrons produced by
highly relativistic particles.
The TOF measures the time of flight of particles 
with a 
100 ps timing resolution.
The ECL detects photons and the deposited energy 
of particles with a
resolution of $\sigma_E/E=1.5\%$ $(2.0\%)$ at 1 GeV (0.1 GeV).  

In processes with quasi-real two-photon collisions
$e^+e^-$ $\to$ $e^+e^-\gamma^*\gamma^*$ $\to$ 
$e^+e^-$ $\Lambda\bar{\Lambda}(\Sigma^0\bar{\Sigma^0})$,
the scattered electrons go down the beam pipe 
and thus only the produced hyperon-pair can be detected. 
The $\gamma\gamma$ axis can be approximated by 
the beam direction in the $e^+e^-$ c.m. frame.
Taking into account the fact that $\Lambda(\bar{\Lambda})$ can be 
reconstructed from its $p\pi^-(\pi^+\bar{p})$ decay products 
and $\Sigma^0(\bar{\Sigma^0})$ from 
$\Lambda\gamma(\bar{\Lambda}\gamma)$,
candidate events including $p\pi^-\pi^+\bar{p}$ candidates
are searched for in a low-multiplicity data stream, where exactly 
four tracks, two positively and two negatively charged, are required.  
The sum of the magnitudes of the momenta of all tracks and the total 
ECL energy are restricted to be below 6 GeV/$c$ and 6 GeV, respectively. 
Each of the tracks satisfies
the first level conditions: 
$p_t>0.1$ GeV/$c$, $dr<5$ cm and $|dz|<5$ cm,
or two or more tracks satisfy the second level conditions:
$p_t>0.3$ GeV/$c$, $dr<1$ cm, $|dz|<5$ cm and 
$17^{\circ}<\theta<150^{\circ}$. Here $p_t$ is the transverse momentum
with respect to the beam axis, 
$dr$ and $dz$ are the radial and axial coordinates of the point 
of closest approach to the nominal collision point respectively,
and $\theta$ is the polar angle of momentum with respect to the electron
beam. Among the first level tracks, 
two tracks with opposite charge are required;  
using the second level tracks,
the invariant mass and the missing mass squared (with a zero mass 
assumption) have to be smaller than 4.5 GeV/$c^2$ and larger than
2 GeV$^2/c^4$, respectively. 

\begin{figure}[htb]
\includegraphics[width=0.49\textwidth]{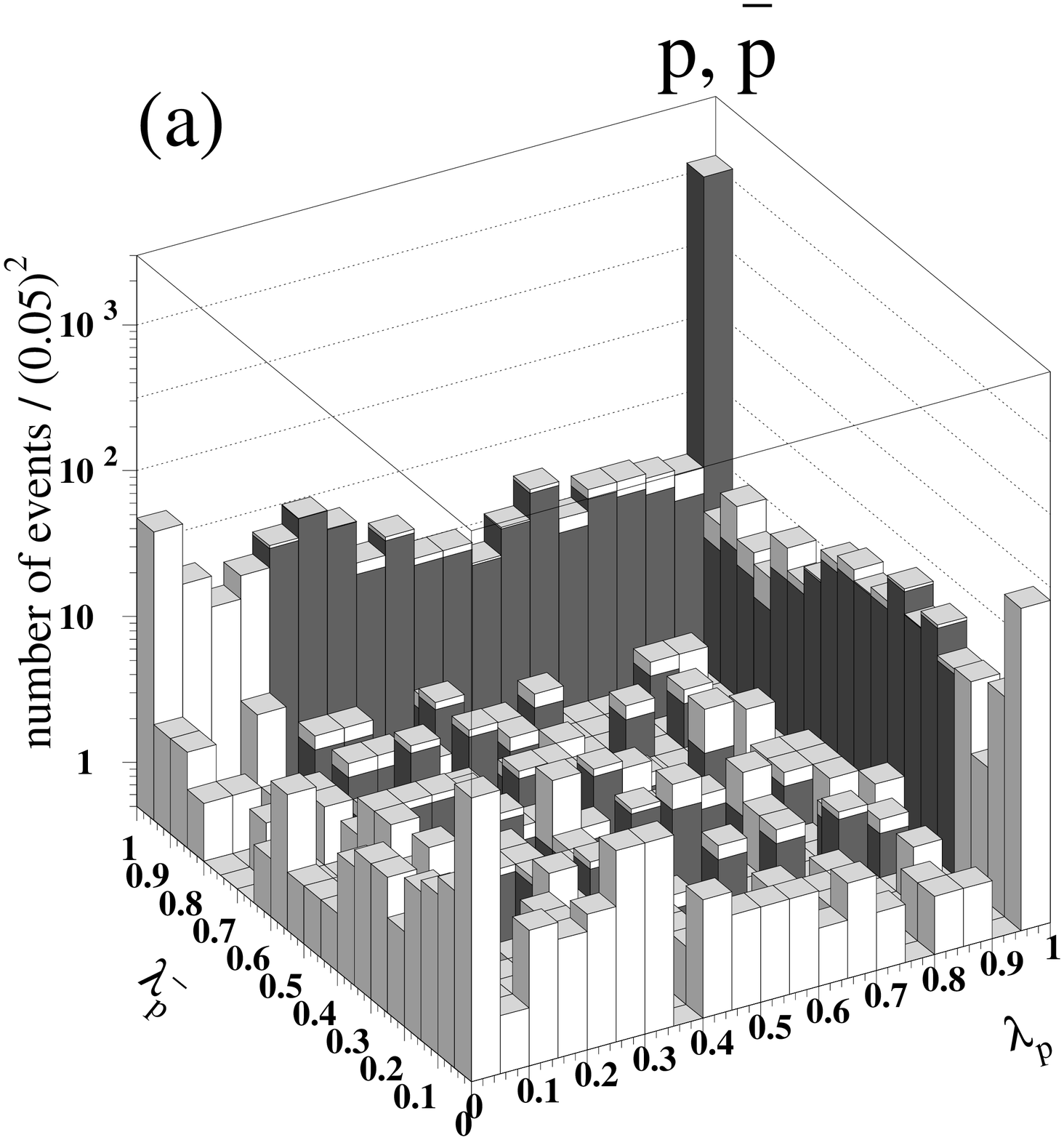}
\includegraphics[width=0.49\textwidth]{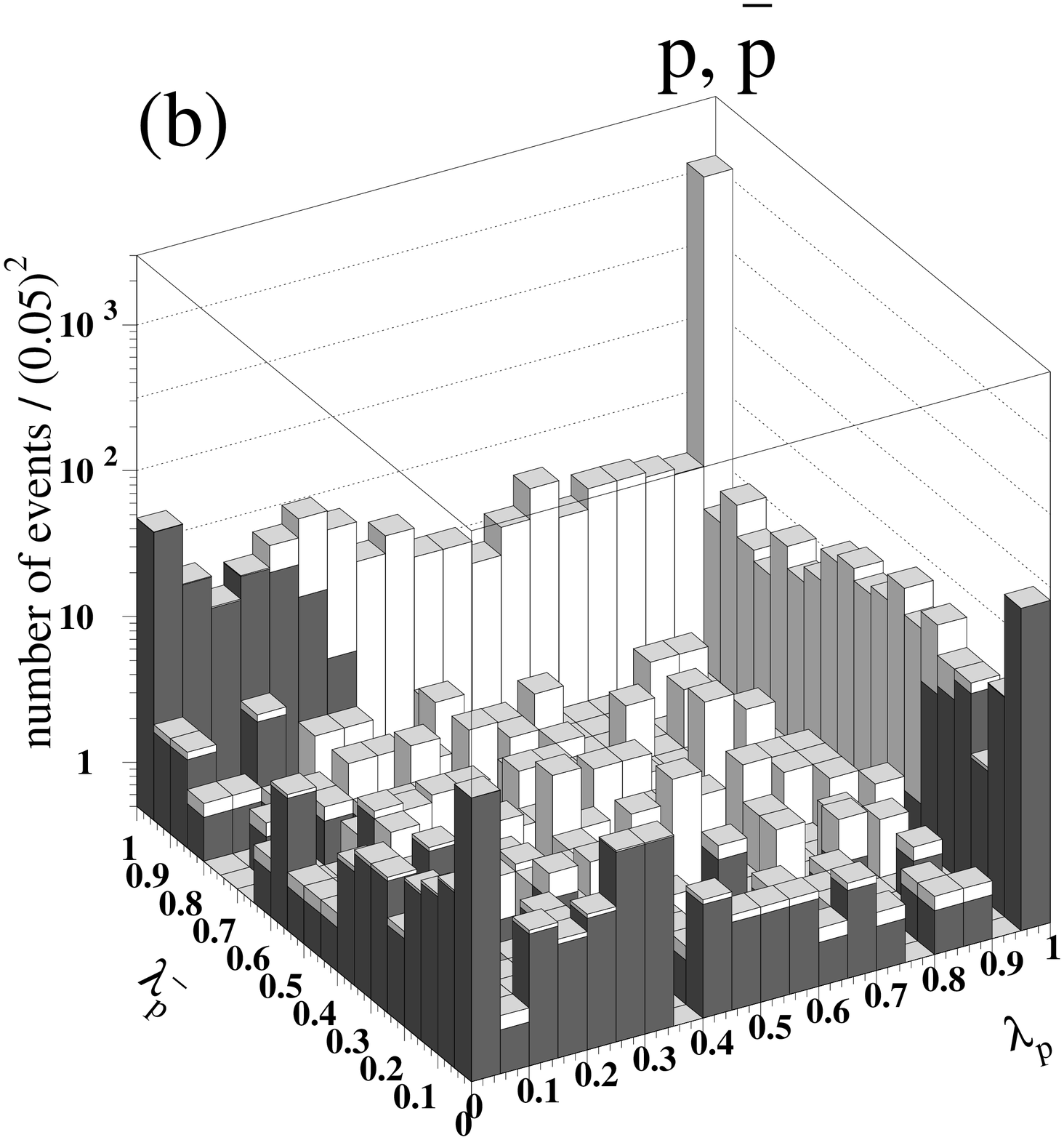}
\vspace{-1.cm}
\caption{Two-dimensional distribution of the normalized likelihood of
the $p$ and the $\bar{p}$ candidates, for the events passing all 
selection criteria up to $|\Sigma p_t^*(\Lambda\bar{\Lambda})|<0.2$ GeV/$c$
except the cuts on the normalized likelihood. In (a), 
the dark bins show the events passing the cuts on the 
normalized likelihood; in (b), the dark bins show the events with either of the 
two tracks satisfying $L_p/L_x<0.5$, where $x$ is $K$, $\pi$, $\mu$ 
or $e$.}
\end{figure}

\begin{figure}[tb]
\includegraphics[width=0.46\textwidth]{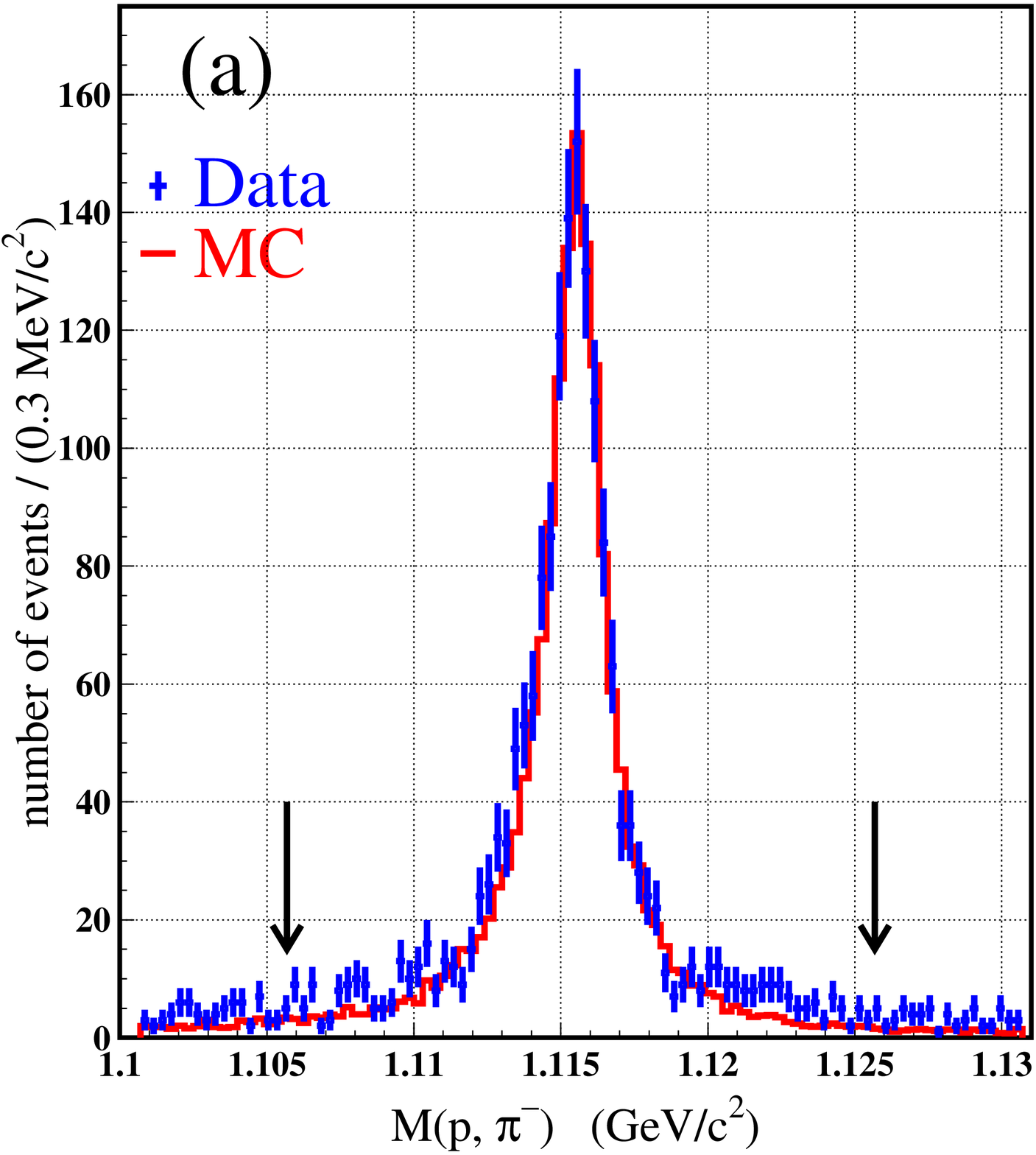}
\includegraphics[width=0.46\textwidth]{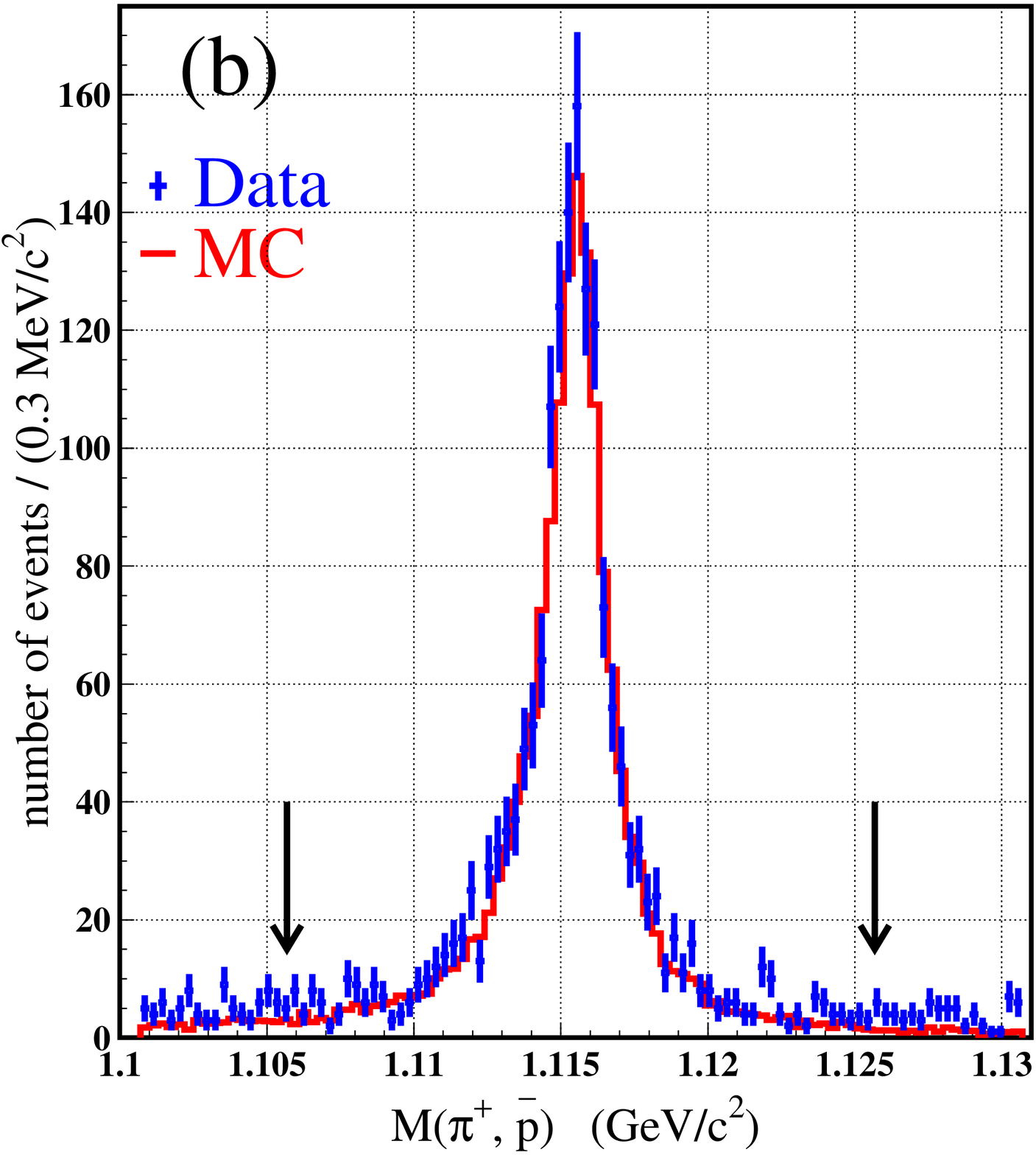}
\includegraphics[width=0.46\textwidth]{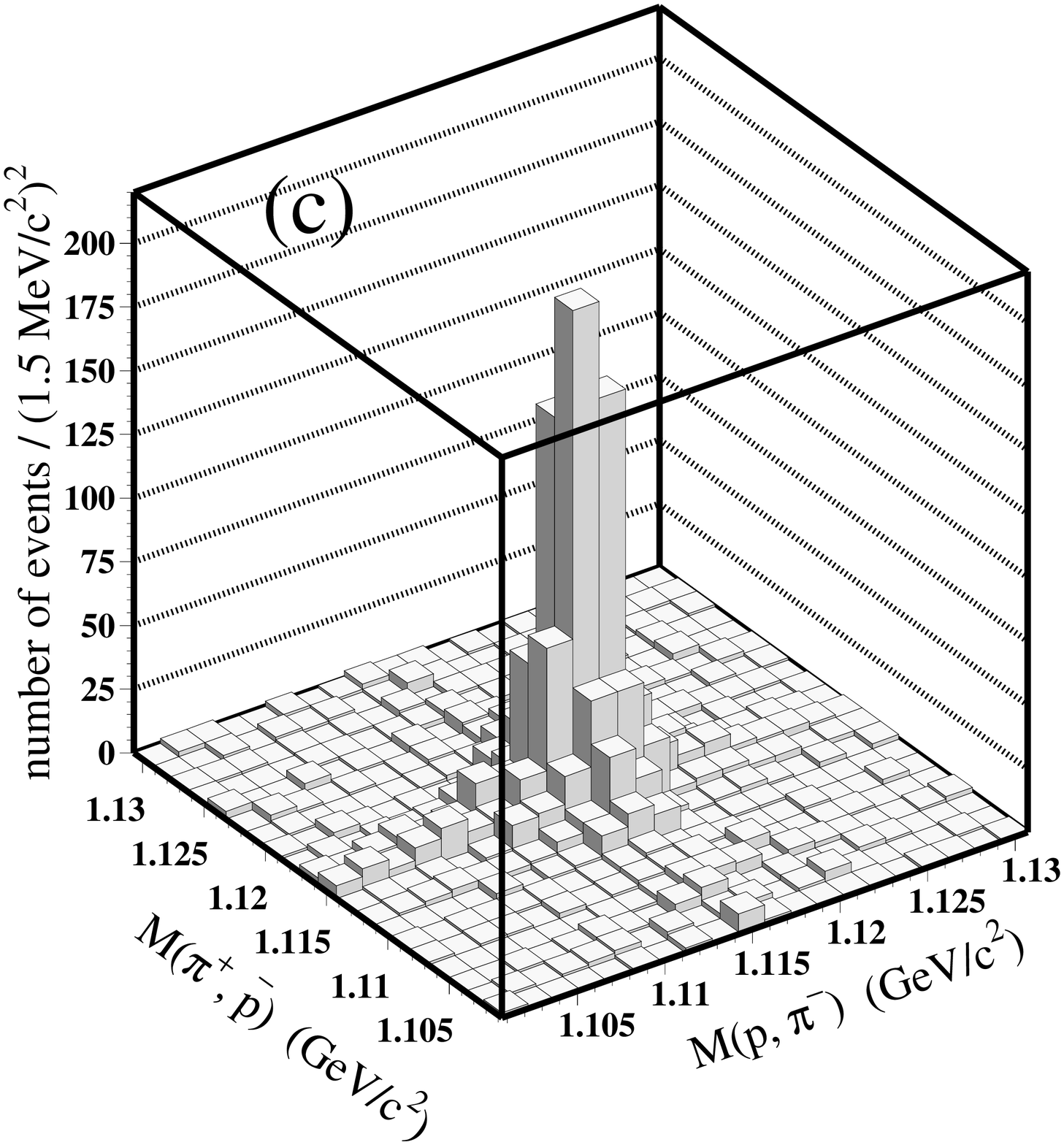}
\vspace{-0.6cm}
\caption{Distributions of the invariant mass of (a) $p,\pi^-$ and (b) $\pi^+,\bar{p}$ 
for the events passing all selection criteria up to 
$|\Sigma p_t^*(\Lambda\bar{\Lambda})|<0.2$ GeV/$c$
except the cut $|\Delta m_{\Lambda(\bar{\Lambda})}|<10$ MeV/$c^2$, 
indicated by the arrows. The Monte Carlo is scaled by a factor corresponding to the  
normalization to data in the $p,\pi^-$ invariant mass range
$1114.8-1116.3$ MeV/$c^2$.  
The two-dimensional distribution is shown in (c).}
\end{figure}

Backgrounds are further reduced by a particle identification (PID) algorithm; 
two tracks with opposite charge have to pass the $p(\bar{p})$ 
identification, and the other two have to pass the $\pi^{\pm}$ identification.
The $p(\bar{p})$ identification 
includes the following conditions \cite{ppbar}:
\begin{itemize}
\item[1.] the difference between the measured and the expected CDC $dE/dx$  
is less than 6 times the resolution:
\begin{equation}
\chi^2_{dE/dx}\equiv 
\bigg[\frac{\Delta(dE/dx)}{\sigma_{dE/dx}}\bigg]^2 <6^2;   
\end{equation}
\item[2.] the ratio of the associated ECL energy to the momentum
is less than 0.9 for the positively charged track;
\item[3.] the number of the photoelectrons in the ACC associated with the 
track is less than 10;
\item[4.] the likelihoods for each particle assigment are combined to
determine the normalized likelihood, 
\begin{equation}
\lambda_p\equiv\frac{L_p}{L_p+L_K+L_{\pi}+L_{\mu}+L_e},
\end{equation}
which has to be larger than 0.95 (0.2) in case an associated TOF hit
is available (not available).   
Each likelihood 
\begin{equation}
L\equiv \mathrm{exp}\bigg[-\frac{1}{2}(\chi^2_{dE/dx}+\chi^2_T)\bigg] 
\end{equation}
is calculated 
using CDC ($dE/dx$) and TOF (time of flight $T$) information \cite{ppbar}.
\end{itemize}
Figure 1 shows the two-dimensional distribution of the normalized
likelihood for the $p,\bar{p}$ candidates before the cuts on 
$\lambda_p$,
where the dark region in (a) shows the events passing the cuts,
and that in (b) shows the events with at least one of the
tracks satisfying $L_p/L_x<0.5$ with $x$ being $K$, $\pi$, $\mu$ or
$e$, which is referred to as the backgound-dominant region and is 
almost rejected.

For $\pi^{\pm}$ identification, only $dE/dx$ is used; 
the pion candidates must have $dE/dx$ consistent with the pion
hypothesis ($|dE/dx-(dE/dx)_{\pi^{\pm}}| < 6~ \sigma_{dE/dx}$)
 and satisfy a proton veto
($|dE/dx-(dE/dx)_{p(\bar{p})}|>6~ \sigma_{dE/dx}$).


The tracks passing the above PID conditions are paired into 
$p\pi^-(\pi^+\bar{p})$ combinations, and each of them has to 
pass the $\Lambda(\bar{\Lambda})$ vertex reconstruction.
The difference between the invariant mass of $p\pi^-(\pi^+\bar{p})$
and the $\Lambda(\bar{\Lambda})$ mass, $\Delta m_{\Lambda(\bar{\Lambda})}$, 
is required to be less than 10 MeV/$c^2$ (Fig.~2).

\vspace{-0.3cm}
\begin{figure}[htb]
\includegraphics[width=0.49\textwidth]{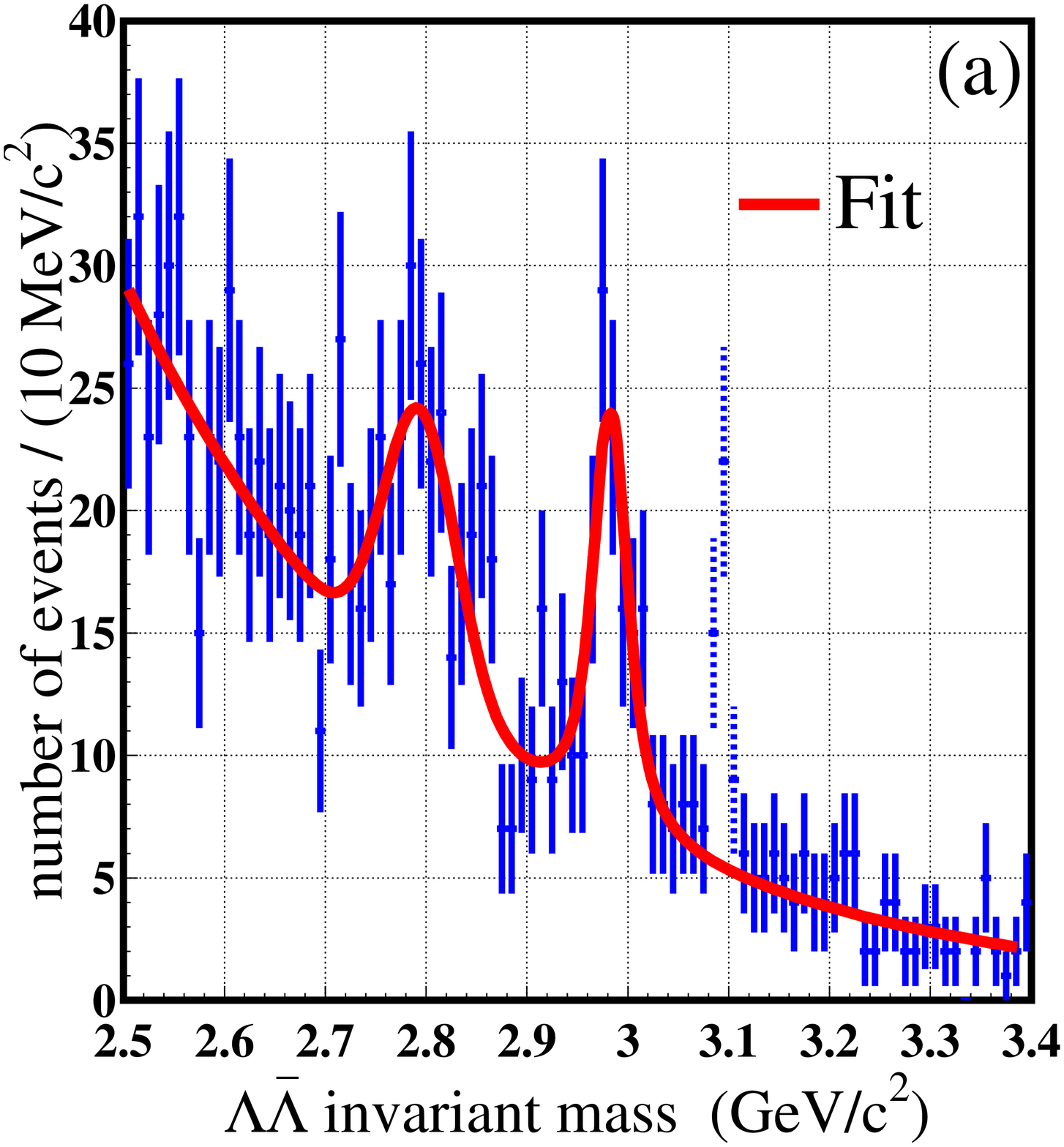}
\includegraphics[width=0.49\textwidth]{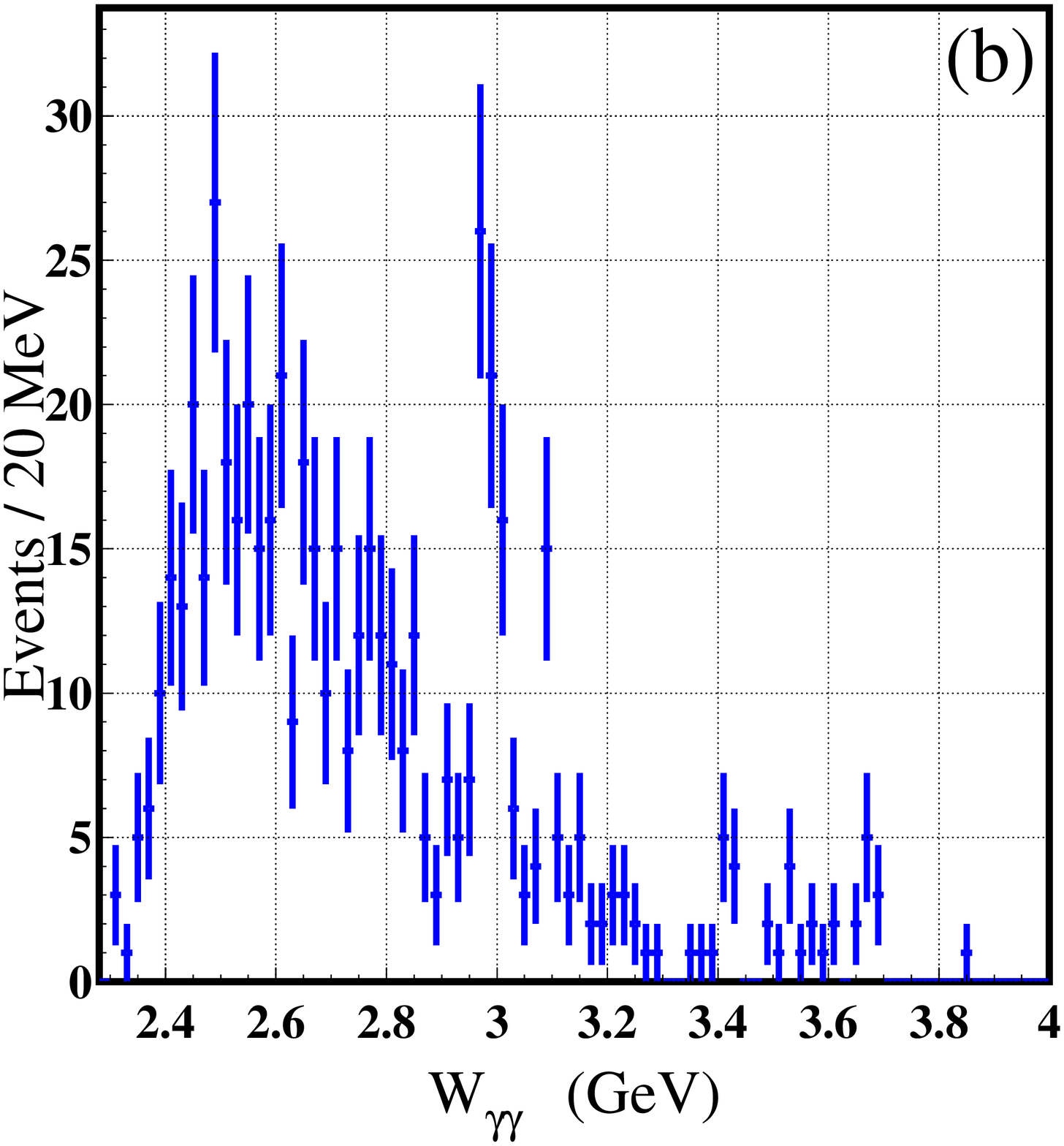}
\vspace{-0.8cm}
\caption{Distribution of $\Lambda\bar{\Lambda}$ invariant mass for 
(a) the events passing the selection criteria up to 
$|\Sigma p_t^*(\Lambda\bar{\Lambda})|<0.2$ GeV/$c$ in the 
$2.5-3.4$ GeV/$c^2$ mass region, and for (b) the events passing the 
exclusive $\gamma\gamma\to\Lambda\bar{\Lambda}$ selection with
$|\Sigma p_t^*(\Lambda\bar{\Lambda})|<0.05$ GeV/$c$. In (a),
the data in the mass range $3.08-3.11$ GeV/$c^2$, corresponding to the region of
radiative return to $J/\psi$ background, are not used in the fit.
}
\end{figure}

\begin{figure}[tb]
\includegraphics[width=0.46\textwidth]{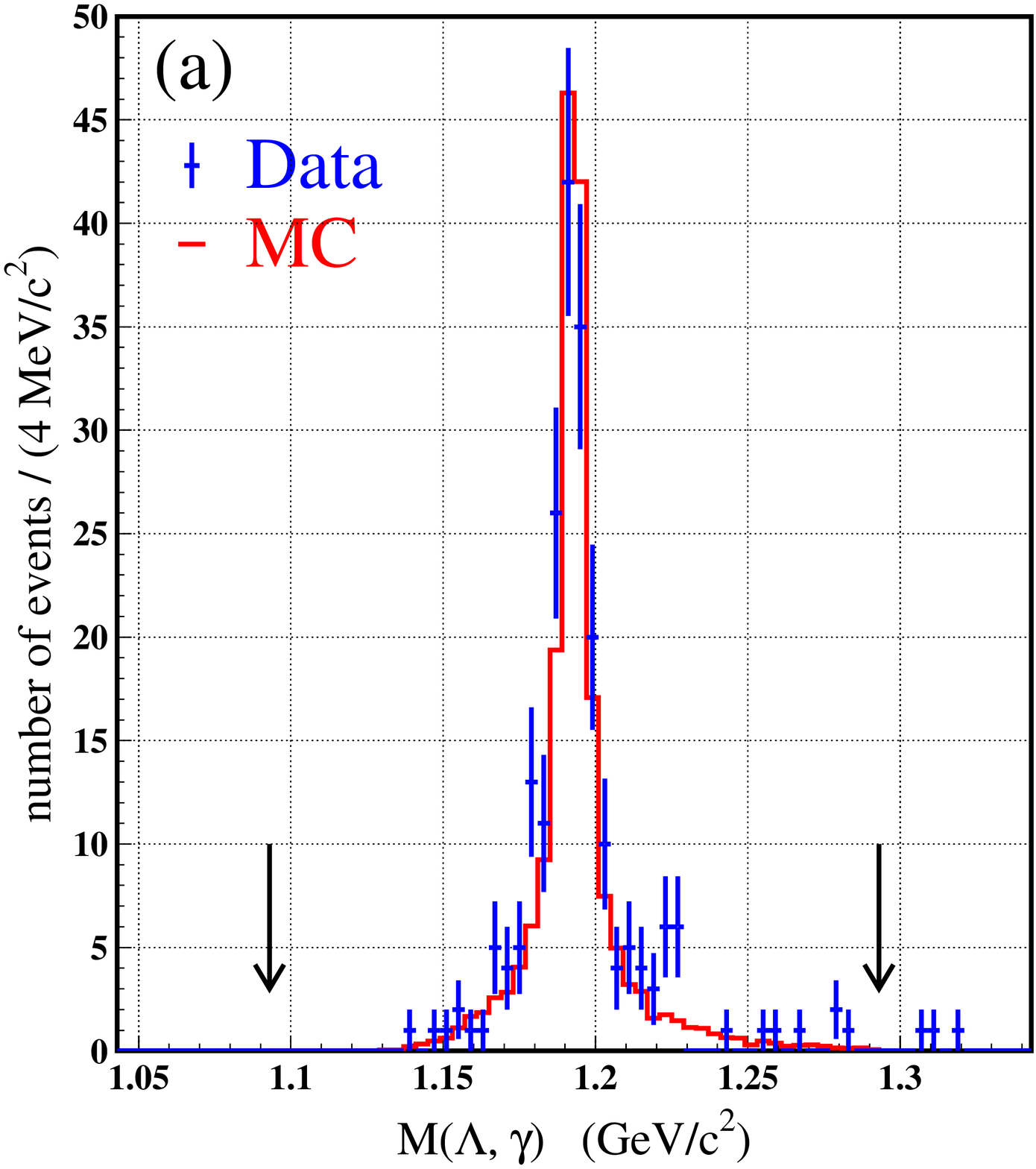}
\includegraphics[width=0.46\textwidth]{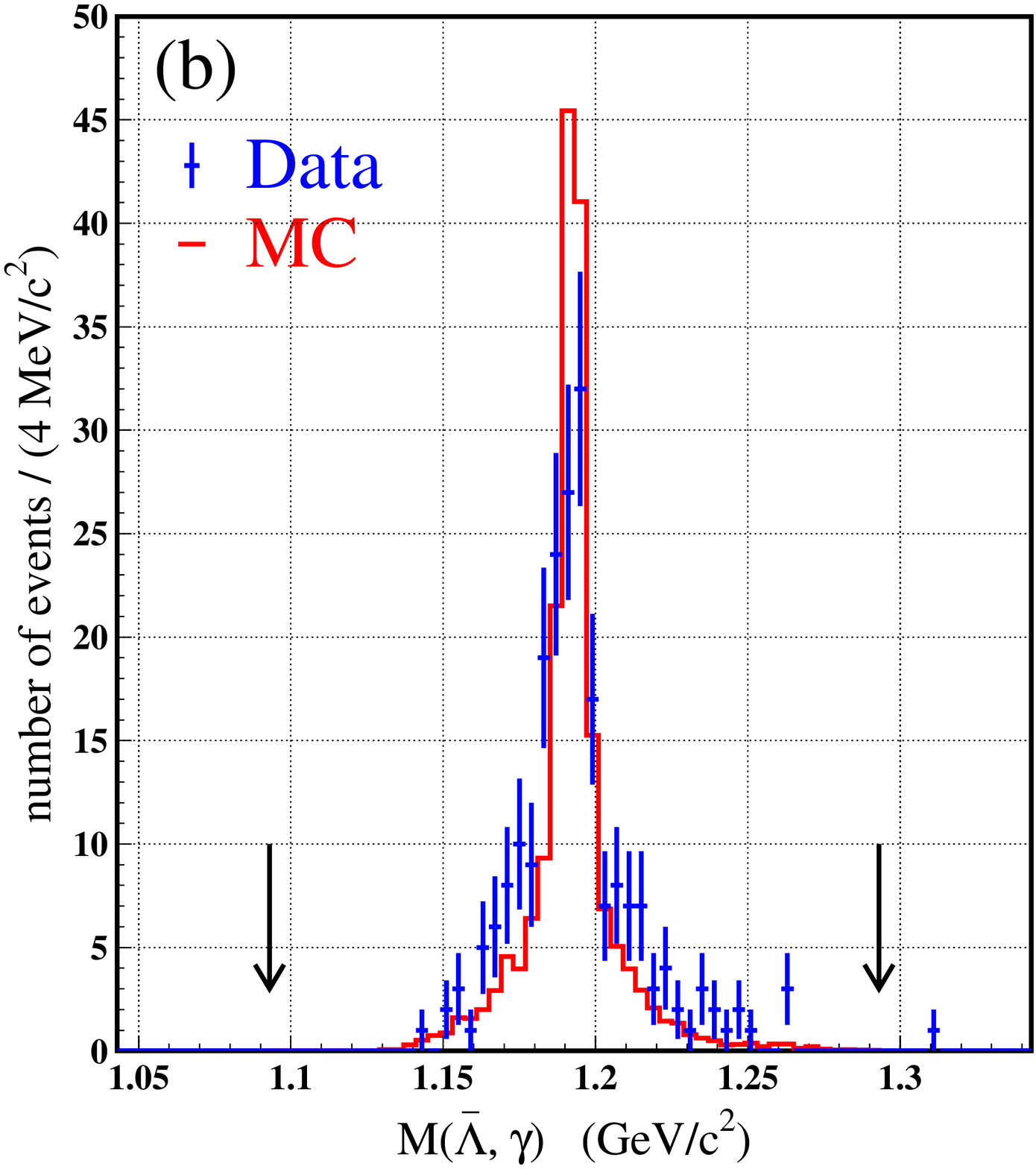}
\includegraphics[width=0.46\textwidth]{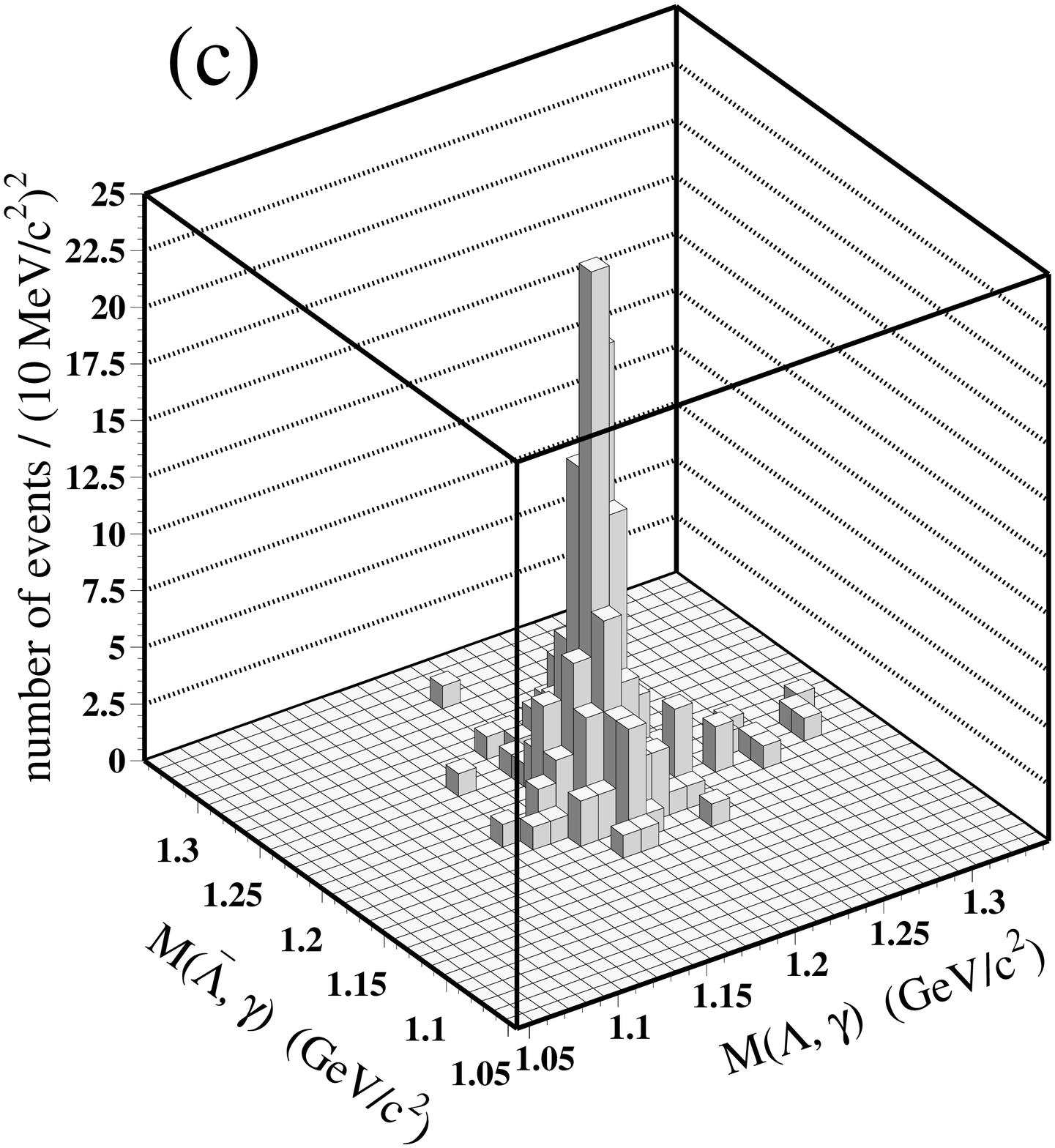}
\vspace{-0.6cm}
\caption{Distributions of the invariant mass of (a) $\Lambda,\gamma$ and (b) 
$\bar{\Lambda},\gamma$ for the events passing all selection criteria for
exclusive $\gamma\gamma\to\Sigma^0\bar{\Sigma^0}$  
except the cut $|\Delta m_{\Sigma^0(\bar{\Sigma^0})}|<100$ MeV/$c^2$ 
indicated by the arrows. The Monte Carlo includes the backgrounds from
$\gamma\gamma\to\Lambda\bar{\Lambda},\Lambda\bar{\Lambda}\pi^0,
\Sigma^0\bar{\Sigma^0}\pi^0$
and is scaled by a factor corresponding to 
the normalization to data in the $\Lambda,\gamma$ invariant mass range
$1181-1201$ MeV/$c^2$.  
The two-dimensional distribution is shown in (c).}
\end{figure}

In order to reject most of the non-exclusive
$\gamma\gamma\to\Lambda\bar{\Lambda}X,\Sigma^0\bar{\Sigma^0}X$ backgrounds,
the transverse momentum balance in the c.m. frame of $e^+e^-$ beams
is required to satisfy
\begin{equation}
|\Sigma p_t^*(\Lambda\bar{\Lambda})|\equiv 
\Bigg|\sum_{i=1}^4\vec{p}_{t_i}^{\hspace{0.1cm}*}\Bigg|
<0.2\hspace{0.2cm}\mathrm{GeV}/c,
\end{equation}
where $\vec{p}_{t_i}^{\hspace{0.1cm}*}$ denotes the transverse momentum
of each track in the $e^+e^-$ c.m. frame. 
Disregarding the two photons from the  
$\Sigma^0\bar{\Sigma^0}\to\Lambda\gamma\bar{\Lambda}\gamma$ decay,
the $|\Sigma p_t^*(\Lambda\bar{\Lambda})|$ 
peak of the $\gamma\gamma\to\Sigma^0\bar{\Sigma^0}$ events 
is around $0.1$ GeV/$c$ 
according to Monte Carlo simulation,
so that most of them 
pass the condition in Eq.~(4).  
In total, 1628 events survive after all of the selection criteria above.
At this stage, peaks around 2.98 and 2.80 GeV/$c^2$ due to the 
intermediate $\eta_c$ resonance, corresponding to the process 
$\gamma\gamma\to\eta_c\to\Lambda\bar{\Lambda},\Sigma^0\bar{\Sigma^0}$,
are observed in the distribution of $\Lambda\bar{\Lambda}$ invariant
mass (Fig.~3(a)). Details will be given in the next section.  
Exclusive $\gamma\gamma\to\Lambda\bar{\Lambda}$ events are further 
required to satisfy $|\Sigma p_t^*(\Lambda\bar{\Lambda})|<0.05$ GeV/$c$,
where 549 events are obtained (Fig.~3(b)).

In the sample of 1628 events satisfying 
$|\Sigma p_t^*(\Lambda\bar{\Lambda})|<0.2$ GeV/$c$, exclusive
$\gamma\gamma\to\Sigma^0\bar{\Sigma^0}$ events are selected as follows. 
Each $\Lambda(\bar{\Lambda})$ is paired with a photon candidate, where
photons are selected from the ECL clusters satisfying
the following conditions: 
\begin{itemize}
\item[1.] the cluster is not associated to any CDC track;
\item[2.] the total energy of the cluster is between 
50 and 200 MeV;
\item[3.] the cluster is photon-like and isolated: $E_9/E_{25}>0.9$,
where $E_9$ and $E_{25}$ are the energy deposition in the $3\times 3$ and 
$5\times 5$ matrix around the ECL crystal with the maximum energy. 
\end{itemize}
The $\Lambda\gamma(\bar{\Lambda}\gamma)$ pair has to satisfy 
$|\Delta m_{\Sigma^0(\bar{\Sigma^0})}|<100$ MeV/$c^2$,
and among all possible combinations the one with the smallest  
$|\Delta m_{\Sigma^0(\bar{\Sigma^0})}|$ is selected (Fig.~4).
Here $|\Delta m_{\Sigma^0(\bar{\Sigma^0})}|$ is 
the difference between the 
invariant mass of $\Lambda\gamma(\bar{\Lambda}\gamma)$ and the 
$\Sigma^0(\bar{\Sigma^0})$ mass.   
 
\vspace{-3.8cm}
\begin{figure}[htb]
\includegraphics[width=0.85\textwidth]{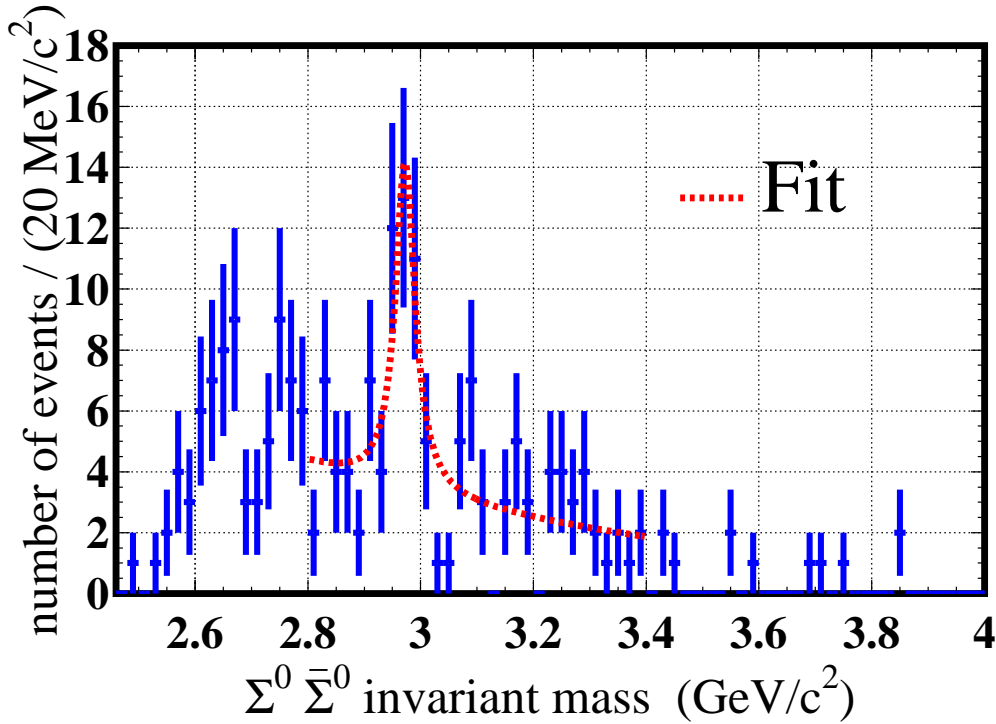}
\vspace{-3.cm}
\caption{Distribution of $\Sigma^0\bar{\Sigma^0}$ 
invariant mass for the events passing the exclusive
$\gamma\gamma\to\Sigma^0\bar{\Sigma^0}$ selection requirements.}
\end{figure}

In order to remove the background from exclusive 
$\gamma\gamma\to\Lambda\bar{\Lambda}$ events, events with
$|\Sigma p_t^*(\Lambda\bar{\Lambda})|<0.06$ GeV/$c^2$ and
$|\Sigma p_t^*(\Lambda\gamma\bar{\Lambda}\gamma)|>0.02$ GeV/$c^2$
are excluded, where $|\Sigma p_t^*(\Lambda\gamma\bar{\Lambda}\gamma)|$
is the transverse momentum balance including the contribution from
the two photons. Furthermore, in order to reject other non-exclusive
backgrounds, $|\Sigma p_t^*(\Lambda\gamma\bar{\Lambda}\gamma)|$ and 
$|\Sigma p_t^*(\Lambda\bar{\Lambda})|$ are restricted 
to below 0.14 and 0.16 GeV/$c^2$, respectively.
About $11\%$ of the events have two or more photon candidates, but   
the $\Lambda$ and $\bar{\Lambda}$ are not paired with two different 
photons; these events are simply omitted.  
After all the above requirements, 
212 events remain (Fig.~5).  

\section{Observation of 
$\gamma\gamma\to\eta_c\to\Lambda\bar{\Lambda},\Sigma^0\bar{\Sigma^0}$}

From the sample of the 1628 events satisfying 
$|\Sigma p_t^*(\Lambda\bar{\Lambda})|<0.2$ GeV/$c$,
peaks around 2.98 and 2.80 GeV/$c^2$ in the distribution of 
the $\Lambda\bar{\Lambda}$ invariant mass (Fig.~3(a)) are identified as 
decays of the $\eta_c(1S)$ resonance \cite{PDG}. The former corresponds
to the $\gamma\gamma\to\eta_c\to\Lambda\bar{\Lambda}$ signal; 
while based on a Monte Carlo study (Fig.~6), 
the latter is identified as the effective signal from
$\gamma\gamma\to\eta_c\to\Sigma^0\bar{\Sigma^0}\to
\Lambda\gamma\bar{\Lambda}\gamma$,
where
the two photons in the final state are not detected. When the 
exclusive $\Sigma^0\bar{\Sigma^0}$ selection requirements are applied,
the latter signal is recovered at the mominal $\eta_c(1S)$ mass 
in the $\Lambda\gamma\bar{\Lambda}\gamma$
invariant mass distribution (Fig.~5).

\begin{figure}[htb]
\includegraphics[width=0.54\textwidth]{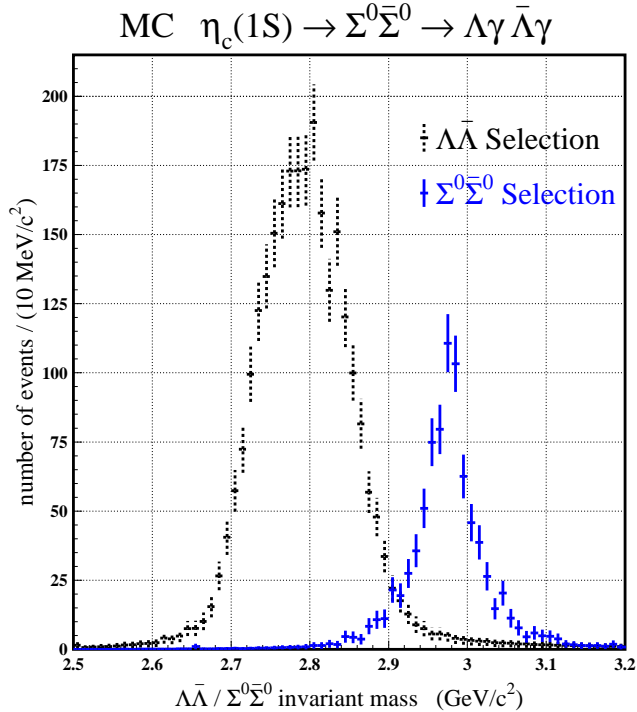}
\vspace{-1.cm}
\caption{Distribution of $\Lambda\bar{\Lambda}(\Sigma^0\bar{\Sigma^0})$ 
invariant mass from a Monte Carlo 
$\gamma\gamma\to\eta_c\to\Sigma^0\bar{\Sigma^0}$ sample passing the 
$\Lambda\bar{\Lambda}(\Sigma^0\bar{\Sigma^0})$ 
selection, respectively.}
\end{figure}

The $\eta_c\to\Lambda\bar{\Lambda}$ signal in Fig.~3(a)
is fitted with a Breit-Wigner function, and an $\eta_c$ yield
of $101.2\pm16.4^{+1.2}_{-3.0}$ events is obtained. The Breit-Wigner 
function for the $\eta_c$ signal is smeared with detector resolution 
determined by Monte Carlo simulation, and the mass and the width of
the $\eta_c$ are also free parameters in the fit. The $\eta_c$
mass and width obtained from the fit are consistent with 
those reported in Ref.~\cite{PDG}. Backgrounds are 
fitted with a smooth function (exponential of a first-order polynomial) 
for the continuum part and a Gaussian for the  
$\eta_c\to\Sigma^0\bar{\Sigma^0}$ contribution. For the latter,
the Gaussian mean value and the width are also free parameters in the fit. 
We obtain a statistical significance of $6.6\sigma$ for the 
$\eta_c\to\Lambda\bar{\Lambda}$ signal from the fit.
Here the statistical significance is defined as 
$\sqrt{-2~\mathrm{ln}(L_0/L_S)}$,
where $L_S$ and $L_0$ denote the maximum likelihoods
of the fits with and without a signal component, respectively.
A similar fit is performed for the $\eta_c\to\Sigma^0\bar{\Sigma^0}$
signal in Fig.~5, using the same type of functions as above for the 
signal and the continuum background. 
An $\eta_c$ signal of $36.1\pm9.2^{+0.0}_{-1.2}$ events 
with a statistical significance of 
$3.9\sigma$ is obtained.
For the systematic error in the $\eta_c$ yield, effects from the 
uncertainties of the continuum background shape and the energy 
detection resolution of signal are taken into account.

The $\eta_c$ yield ($N_{\eta_c}$) is converted to 
a product of the two-photon width of $\eta_c$ and its 
branching fractions to $\Lambda\bar{\Lambda}$ or $\Sigma^0\bar{\Sigma^0}$, 
according to the formula 
\begin{equation}
\Gamma_{\gamma\gamma}(\eta_c)\times 
B(\eta_c\to\Lambda\bar{\Lambda},\Sigma^0\bar{\Sigma^0}) 
=\frac{N_{\eta_c} m_{\eta_c}^2}{ 
4\pi^2\hspace{0.08cm}D\hspace{0.08cm}\varepsilon\hspace{0.08cm}
L_{\rm int}\hspace{0.08cm}
dL_{\gamma\gamma}/dW_{\gamma\gamma}
} 
\end{equation}
using the luminosity function
$dL_{\gamma\gamma}/dW_{\gamma\gamma}$ (Eq.~(7)) determined at the energy of
the $\eta_c$ mass ($m_{\eta_c}$) with efficiency $\varepsilon$ from 
Monte Carlo. Here $D$ denotes the branching fraction for 
$\Lambda\bar{\Lambda}\to p\pi^-\pi^+\bar{p}$, which is equal to
$40.83\pm 0.64\%$ \cite{PDG},
and $L_{\mathrm{int}}$ denotes the integrated luminosity
which is equal to 464 fb$^{-1}$. The results are shown in Table I,
where the systematic errors are estimated taking into account
the sources listed in Table II.
The value of $B(\eta_c\to\Lambda\bar{\Lambda})$ given in Table I 
is in agreement with the Belle measurement from $B$ meson decays  
\cite{MZW}.
 
\begin{table}[htb]
\caption{Measured $\Gamma_{\gamma\gamma}(\eta_c)\times
B(\eta_c\to\Lambda\bar{\Lambda},\Sigma^0\bar{\Sigma^0})$ and 
$B(\eta_c\to\Lambda\bar{\Lambda},\Sigma^0\bar{\Sigma^0})$. The latter is 
obtained using $\Gamma_{\gamma\gamma}(\eta_c)=7.0^{+1.0}_{-0.9}$ keV 
\cite{PDG}. The errors are statistical and systematic respectively. 
The last error in $B(\eta_c\to\Lambda\bar{\Lambda},\Sigma^0\bar{\Sigma^0})$ 
is from the uncertainty in $\Gamma_{\gamma\gamma}(\eta_c)$.
}
\begin{tabular}{@{\hspace{0.5cm}}c@{\hspace{0.5cm}}||@{\hspace{0.5cm}}c@{\hspace{0.5cm}}||@{\hspace{0.5cm}}c@{\hspace{0.5cm}}}
\hline\hline 
$\eta_c$ decay mode & 
$\Gamma_{\gamma\gamma}(\eta_c)\times 
B(\eta_c\to\Lambda\bar{\Lambda},\Sigma^0\bar{\Sigma^0})$ (eV) & 
$B(\eta_c\to\Lambda\bar{\Lambda},\Sigma^0\bar{\Sigma^0})\times 10^3$ \\
\hline
 $\eta_c\to\Lambda\bar{\Lambda}$ & 6.21 $\pm$ 1.01 $^{+0.49}_{-0.52}$ & 
0.89 $\pm$ 0.14 $^{+0.07}_{-0.07}$ $^{+0.13}_{-0.11}$  \\ 
 $\eta_c\to\Sigma^0\bar{\Sigma^0}$ & 9.80 $\pm$ 2.50 $^{+0.98}_{-1.03}$ & 
1.40 $\pm$ 0.36 $^{+0.14}_{-0.15}$ $^{+0.20}_{-0.18}$  \\ 
\hline\hline
\end{tabular}
\end{table}

\begin{table}[htb]
\caption{Systematic errors ($\%$) for the measured 
$\Gamma_{\gamma\gamma}(\eta_c)\times 
B(\eta_c\to\Lambda\bar{\Lambda},\Sigma^0\bar{\Sigma^0})$ and 
$B(\eta_c\to\Lambda\bar{\Lambda},\Sigma^0\bar{\Sigma^0})$
}
\begin{tabular}{@{\hspace{0.4cm}}l@{\hspace{0.4cm}}||@{\hspace{0.4cm}}c@{\hspace{0.4cm}}||@{\hspace{0.4cm}}c@{\hspace{0.4cm}}}
\hline\hline 
Source &
$\eta_c\to\Lambda\bar{\Lambda}$ & $\eta_c\to\Sigma^0\bar{\Sigma^0}$  
\\
\hline 
 Integrated luminosity &  
 1.4 & 1.4 
\\
 Luminosity function & 
 4 & 4 
\\
 Branching fraction of 
 $\Lambda\bar{\Lambda}\to p\pi^-\pi^+\bar{p}$ &
 1.6 & 1.6 
\\
 Monte Carlo statistics & 
 1.0 & 2.4 
\\
 Trigger efficiency & 
 4.3 & 4.3 
\\
 Particle identification 
 efficiency 
& 
 4.5 & 4.5 
\\
 Photon selection efficiency &
 0 & 5.9 
\\
 Background shape and 
 energy detection resolution of signal
&  
  $^{+1.2}_{-3.0}$ 
& 
  $^{+0.0}_{-3.3}$  
\\
\hline
 Total &   
  $^{+7.9}_{-8.3}$ 
&  
  $^{+10.0}_{-10.5}$  
\\
\hline\hline
\end{tabular}
\end{table}

\section{Measurement of the cross sections for 
$\gamma\gamma\to\Lambda\bar{\Lambda},\Sigma^0\bar{\Sigma^0}$}

The cross sections for 
$\gamma\gamma\to\Lambda\bar{\Lambda}(\Sigma^0\bar{\Sigma^0})$ are measured, 
using the data sample of 549 (212) events (Fig.~3(b) and 5) passing 
the exclusive
$\gamma\gamma\to\Lambda\bar{\Lambda}(\Sigma^0\bar{\Sigma^0})$
selection requirements.
The number of events $\Delta N(W_{\gamma\gamma},|\cos{\theta^*}|)$
and the efficiency 
$\varepsilon(W_{\gamma\gamma},|\cos{\theta^*}|)$ 
are determined for two-dimensional bins of $W_{\gamma\gamma}$ 
and $|\cos{\theta^*}|$, where $W_{\gamma\gamma}$ and $\theta^*$
are the two-photon c.m. energy and the c.m. angle 
of the hyperon, respectively.  
The background
from radiative return to $J/\psi$ 
contributes most of the excess events in the range of 
$W_{\gamma\gamma}=3.08-3.10$ GeV,
and thus the data in that narrow bin are not used.    
The ratio $\Delta N/\varepsilon$ is then converted to 
the differential cross section, according to the formula
\begin{equation}
\frac{d\sigma_{\gamma\gamma\to\Lambda\bar{\Lambda},\Sigma^0\bar{\Sigma^0}}
(W_{\gamma\gamma})}{d|\cos{\theta^*}|}=
\frac{\Delta N/\varepsilon}
{D\hspace{0.15cm}L_{\rm int}\hspace{0.15cm}
\frac{dL_{\gamma\gamma}}{dW_{\gamma\gamma}}\hspace{0.15cm}
\Delta W_{\gamma\gamma}\hspace{0.15cm} 
\Delta |\cos{\theta^*}|}
\end{equation}    
The luminosity function $\frac{dL_{\gamma\gamma}}{dW_{\gamma\gamma}}$,
as a function of $W_{\gamma\gamma}$,
is defined by
\begin{equation}
\sigma_{e^+e^-\to e^+e^-\Lambda\bar{\Lambda}(\Sigma^0\bar{\Sigma^0})}=
\int \sigma_{\gamma\gamma\to\Lambda\bar{\Lambda}(\Sigma^0\bar{\Sigma^0})}
(W_{\gamma\gamma})\hspace{0.25cm}
\frac{dL_{\gamma\gamma}(W_{\gamma\gamma})}{dW_{\gamma\gamma}}\hspace{0.25cm}
dW_{\gamma\gamma} 
\end{equation}
and is calculated by TREPS \cite{TRE} using the equivalent photon
approximation method \cite{BN}. The effects from longitudinal photons 
are neglected. For event generation in TREPS, the
maximum virtuality of each of the two photons, $Q_1^2$ and $Q_2^2$, is
limited to 1 GeV$^2$, and a form factor term is introduced for
the high-$Q^2$ suppression effect,
$(1+Q_1^2/W_{\gamma\gamma}^2)^{-2}(1+Q_2^2/W_{\gamma\gamma}^2)^{-2}$.
The systematic uncertainty in the luminosity function is estimated by
comparing TREPS to a QED calculation including all order $\alpha^4$ 
diagrams \cite{AAF}, and an agreement within $3-5\%$ was reported
for $W_{\gamma\gamma}=2-4$ GeV \cite{TRE,LFE}. 

Overall detection efficiencies from Monte Carlo ($\varepsilon$) are 
obtained from the signal samples generated by the TREPS codes \cite{TRE}.
Other samples of the type
$\gamma\gamma\to\Lambda\bar{\Lambda}\pi^0,\Sigma^0\bar{\Sigma^0}\pi^0$ are 
generated using the GGLU \cite{GGL} code in order to study non-exclusive
backgrounds. Detector simulation is based on GEANT3 \cite{GEA}. 
Typical values of $\varepsilon$ from Monte Carlo simulation
in the region with $|\cos{\theta^*}|<0.1$ 
and $W_{\gamma\gamma}=2.5-4.0$ GeV are 
$\sim 2\%-9\%$ for $\gamma\gamma\to\Lambda\bar{\Lambda}$ and 
$\sim 0.3\%-3.5\%$ for $\gamma\gamma\to\Sigma^0\bar{\Sigma^0}$ (Fig.~7).

\begin{figure}[htb]
\includegraphics[width=0.49\textwidth]{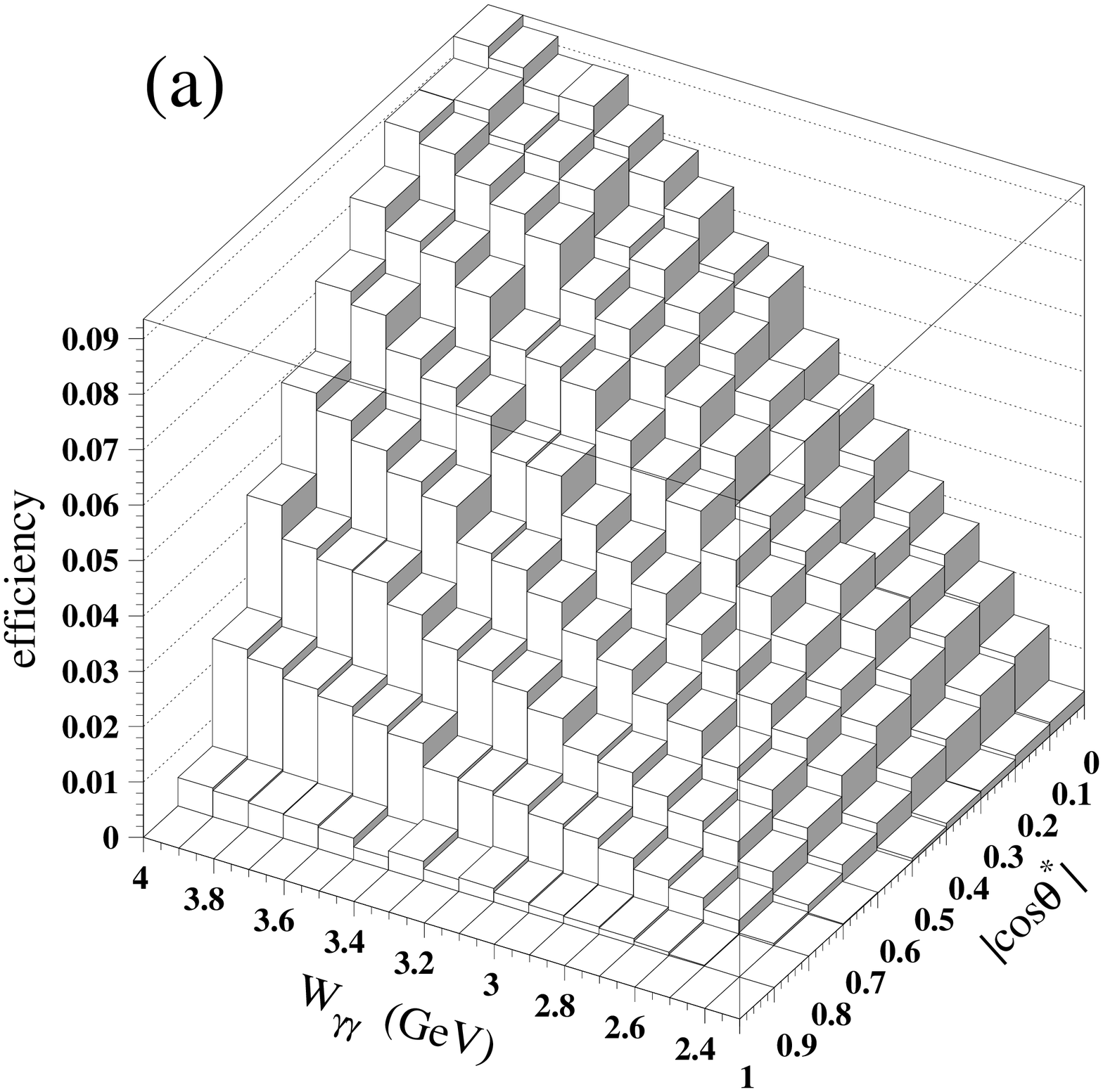}
\includegraphics[width=0.49\textwidth]{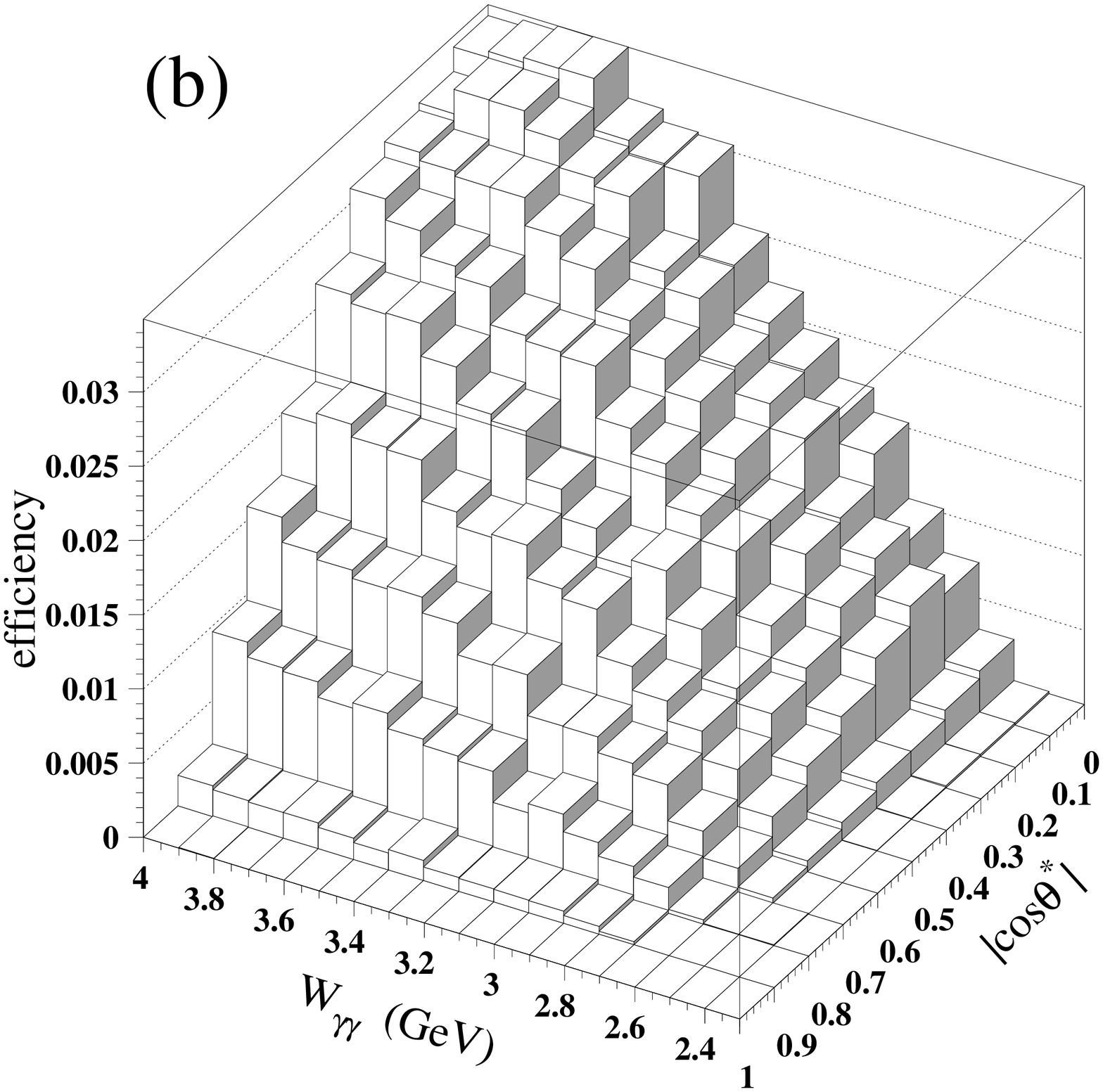}
\vspace{-1.cm}
\caption{Overall detection efficiency as a function of $W_{\gamma\gamma}$
and $|\cos{\theta^*}|$ for (a) $\gamma\gamma\to\Lambda\bar{\Lambda}$ and 
(b) $\gamma\gamma\to\Sigma^0\bar{\Sigma^0}$ events.}
\end{figure}

\begin{figure}[htb]
\includegraphics[width=0.7\textwidth]{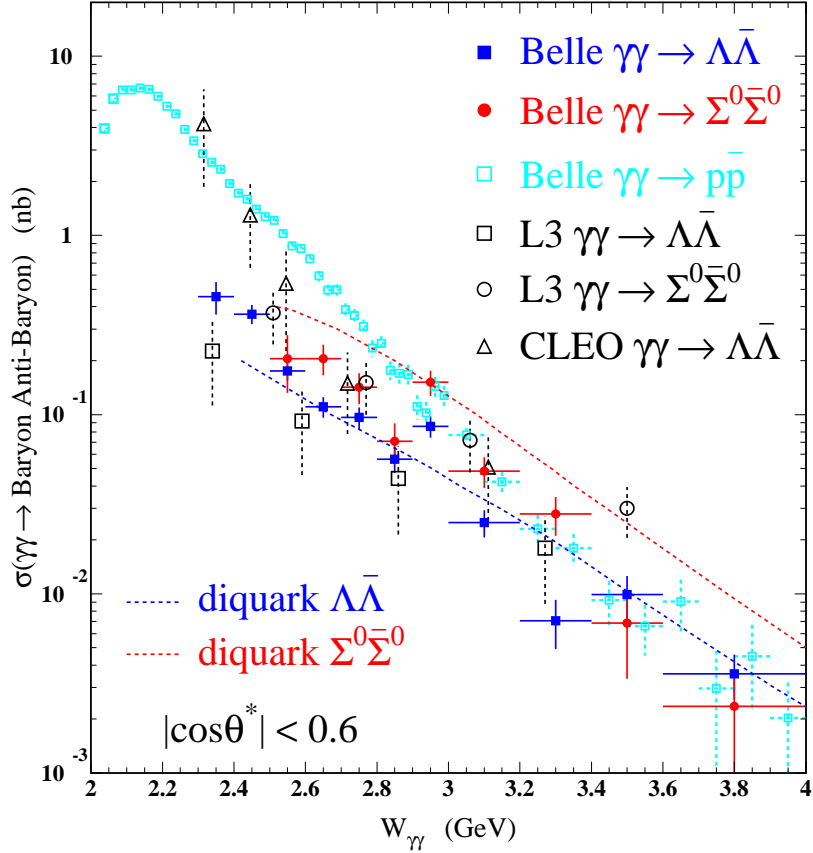}
\vspace{-1.3cm}
\caption{Measured cross sections for $\gamma\gamma\to\Lambda\bar{\Lambda},
\Sigma^0\bar{\Sigma^0},p\bar{p}$.  
The error bars are purely statistical for Belle, and the contribution from the 
intermediate $\eta_c$ decay is included. The theoretical curves shown are
the predictions from the diquark model \cite{CF5}. 
}
\end{figure}

Knowing the differential cross section, the total cross section can be 
obtained by summing over $|\cos{\theta^*}|$:
\begin{equation}
\sigma_{\gamma\gamma\to\Lambda\bar{\Lambda},\Sigma^0\bar{\Sigma^0}}(W_{\gamma\gamma})
=\sum \frac{d\sigma_{\gamma\gamma\to\Lambda\bar{\Lambda},\Sigma^0\bar{\Sigma^0}}
(W_{\gamma\gamma})}
{d|\cos{\theta^*}|}\hspace{0.1cm}  
\Delta |\cos{\theta^*}|
\end{equation}
which is performed with $\Delta|\cos{\theta^*}|=0.1$ 
for $|\cos{\theta^*}|$ up to 0.6. The results are shown in Fig.~8,
where the contribution from the intermediate $\eta_c$ decay is included.
Previous measurements from CLEO \cite{CL}, L3 \cite{L3} and the 
$\gamma\gamma\to p\bar{p}$ cross sections from Belle \cite{ppbar},
as well as the predictions given by the diquark model \cite{CF5}, are also 
shown for comparison. The major sources of systematic error are listed in 
Table III.  

\begin{table}[htb]
\caption{Systematic error ($\%$) for the measured cross sections
for $\gamma\gamma\to\Lambda\bar{\Lambda},\Sigma^0\bar{\Sigma^0}$ 
}
\begin{tabular}{@{\hspace{0.5cm}}l@{\hspace{0.5cm}}||@{\hspace{0.5cm}}c@{\hspace{0.5cm}}||@{\hspace{0.5cm}}c@{\hspace{0.5cm}}}
\hline\hline 
Source & $\gamma\gamma\to\Lambda\bar{\Lambda}$ & $\gamma\gamma\to\Sigma^0\bar{\Sigma^0}$ \\
\hline 
 Integrated luminosity &  
 1.4 & 1.4 \\
 Luminosity function & 
 5 & 5 \\
 Branching fraction of 
 $\Lambda\bar{\Lambda}\to p\pi^-\pi^+\bar{p}$ & 
 1.6 & 1.6 \\
 Monte Carlo statistics & 
 $0.5-8.6$ & $1.0-8.1$ \\
 Trigger efficiency & 
 4.3 & 4.3 \\
 Particle identification 
 efficiency &
 4.5 & 4.5 \\
 Photon selection efficiency &
 0 & 5.9 \\
 Residual background &
 $15-26$ & $10-37$ \\
\hline  
Total & $17-29$ & $14-39$ \\
\hline \hline
\end{tabular}
\end{table}

The Monte Carlo trigger efficiency for the \emph{low energy trigger} is checked using 
experimental data. Over $80\%$ of the signal events pass this 
trigger. The \emph{low energy trigger} requires a 0.5 GeV ECL
total energy sum and at least two 
tracks that go through three inner CDC trigger-layers. 
The trigger 
efficiencies of the two sub-parts of the low energy trigger (i.e. the ECL and the 
CDC parts) are separately obtained using the experimental data passing the 
\emph{two-track trigger} and the \emph{high energy trigger} respectively, 
where the former requires at least
two tracks that go through all CDC layers
and the latter is based on a 1 GeV threshold for an ECL
total energy sum \cite{belle,ECLT}. The overall trigger efficiency is 
$\sim 60-80\%$ corresponding to an average transverse momentum of the 
$\Lambda$ and $\bar\Lambda$ from 0.3 to 1.0 GeV/$c$ data. Corrections for
the Monte Carlo trigger efficiency are implemented according to the data, and the
systematic error is $4.3\%$.  

The systematic uncertainty of the event selection efficiency is predominantly coming
from PID and the photon selection. The latter denotes the 
$\Sigma^0\bar{\Sigma^0}$ selection up to the paring of 
$\Lambda\gamma\bar{\Lambda}\gamma$ to $\Sigma^0\bar{\Sigma^0}$,
which is applied to the events with an identified $\Lambda\bar{\Lambda}$ pair. 
The overall PID and the photon selection efficiencies
are $\sim90\%$ ($\sim86\%$) and $\sim55\%$ ($\sim45\%$) at 
$W_{\gamma\gamma}=3$ GeV (4 GeV). 
The accuracy of the Monte Carlo efficiencies are checked from the 
$\eta_c$ yield in the data,  where a realistic PID
or photon selection efficiency can be estimated from the
number of events with and without the application of the
PID or the photon selection, respectively. 
The systematic errors are $4.5\%$ and $5.9\%$ corresponding to PID and 
the photon selection efficiency respectively, for $W_{\gamma\gamma}$ in the  
$2.7-3.2$ GeV region. 
   
One of the major backgrounds is non-exclusive decays
surviving the transverse momentum balance cut, and the contamination
from such backgrounds is checked from the Monte Carlo samples of 
$\gamma\gamma\to\Lambda\bar{\Lambda}\pi^0,\Sigma^0\bar{\Sigma^0}\pi^0$.
The most conservative estimates are made assuming
these backgrounds are the same size as the signals. 
Another significant residual
background type comes from the signals themselves,
$\gamma\gamma\to\Lambda\bar{\Lambda},\Sigma^0\bar{\Sigma^0}$, which could
contaminate each other according to a Monte Carlo study.
A possible background from 
$\gamma\gamma\to\Lambda\bar{\Sigma^0}(\Sigma^0\bar{\Lambda})$ 
is omitted based on the expected level from current 
theoretical predictions \cite{CF5,HB}.
The backgound from radiative return is at the level of $1\%$ or less.
The overall residual background is $15-26\%$ for
$\gamma\gamma\to\Lambda\bar{\Lambda}$ and is $10-37\%$ for
$\gamma\gamma\to\Sigma^0\bar{\Sigma^0}$, where the ranges are determined
by using the least and the most conservative estimates for the non-exclusive 
background in the data. Since it is difficult to subtract all these possible 
backgrounds accurately, the estimated values above are assigned as     
systematic uncertainties in the measured cross sections.

\section{Conclusion}

Using 464 fb$^{-1}$ of data, cross sections for hyperon pair production 
$\gamma\gamma\to\Lambda\bar{\Lambda},\Sigma^0\bar{\Sigma^0}$ have been
measured at Belle. Intermediate $\eta_c$ resonances are observed, and 
the products of the two-photon width of the $\eta_c$ and its branching
ratios to $\Lambda\bar{\Lambda},\Sigma^0\bar{\Sigma^0}$ are determined.
The result for $\eta_c\to\Sigma^0\bar{\Sigma^0}$ is the first
measurement. 

The measured cross sections for $\gamma\gamma\to\Lambda\bar{\Lambda}$
from Belle are compatible with L3 \cite{L3} results.
The data show that the cross sections for the 
three channels 
$\gamma\gamma\to\Lambda\bar{\Lambda},\Sigma^0\bar{\Sigma^0}, p\bar{p}$ 
converge at high energy. 
The deviation of the theoretical predictions from data
implies that current models have to be improved; the effects 
from flavour symmetry breaking could be non-negligible in the measured 
energy range.

\section{Acknowledgements}

We thank the KEKB group for the excellent operation of the
accelerator, the KEK cryogenics group for the efficient
operation of the solenoid, and the KEK computer group and
the National Institute of Informatics for valuable computing
and Super-SINET network support. We acknowledge support from
the Ministry of Education, Culture, Sports, Science, and
Technology of Japan and the Japan Society for the Promotion
of Science; the Australian Research Council and the
Australian Department of Education, Science and Training;
the National Science Foundation of China and the Knowledge
Innovation Program of the Chinese Academy of Sciencies under
contract No.~10575109 and IHEP-U-503; the Department of
Science and Technology of India; 
the BK21 program of the Ministry of Education of Korea, 
the CHEP SRC program and Basic Research program 
(grant No.~R01-2005-000-10089-0) of the Korea Science and
Engineering Foundation, and the Pure Basic Research Group 
program of the Korea Research Foundation; 
the Polish State Committee for Scientific Research; 
the Ministry of Science and Technology of the Russian
Federation; the Slovenian Research Agency;  the Swiss
National Science Foundation; the National Science Council
and the Ministry of Education of Taiwan; and the U.S.\
Department of Energy.


%

\end{document}